\documentclass[12pt,osajnl,preprint,showpacs]{revtex4}
\begin{document}

\title{Vector modulation instability induced by vacuum
fluctuations in highly birefringent fibers in the anomalous
dispersion regime}

\author{D. Amans}
\author{E. Brainis}
\author{M. Haelterman}
\author{Ph. Emplit}
\affiliation{ Optique et acoustique, Universit\'e Libre de
Bruxelles, avenue F.D. Roosvelt
50, CP 194/5, 1050 Bruxelles, Belgium}%
\author{S. Massar}
\affiliation{Laboratoire d'information quantique and QUIC,
Universit\'{e} Libre de Bruxelles, avenue F.D. Roosvelt
50, CP 165/59, 1050 Bruxelles, Belgium}%

\begin{abstract}
We report a detailed experimental study of vector modulation
instability in highly birefringent optical fibers in the anomalous
dispersion regime. We prove that the observed instability is
mainly induced by vacuum fluctuations. The detuning of the
spectral peaks agrees with linear perturbation analy\-sis. The
exact shape of the spectrum is well reproduced by numerical
integration of stochastic nonlinear Schr\"{o}dinger equations
describing quantum propagation.
\end{abstract}

\maketitle

In birefringent silica fibers, stable propagation of a
monochromatic wave can be inhibited by a nonlinear process called
\emph{vector} modulation instability (V-MI) in both dispersion
regimes (normal or anomalous).\cite{H75,TW92,HH92} This contrasts
with the \emph{scalar} modulation instability (S-MI) that does not
require birefringence but can only arise in the anomalous
dispersion regime (at least when second order dispersion
dominates\cite{PM03,H03}). Two limits, those of weak and strong
birefringence, are amenable to relatively simple analytical
study.\cite{Agrawal95} These predictions have been confirmed
experimentally in a number of
cases,\cite{Rothenberg90,Drummond90,Murdoch95} particularly in the
normal dispersion regime.

The only experimental investigation of V-MI in the anomalous
dispersion regime that we are aware of is a recent unsuccessful
attempt using photonic crystal fibers.\cite{K04} Here we report
what is to our knowledge the first experimental study of V-MI in
strongly birefringent silica fibers in the anomalous dispersion
regime. We also carry out a very precise comparison between
experimental results and the predictions of numerical simulations.

Modulation instabilities (MI) can be induced by classical noise
present initially together with the pump beam. But MI can also
arise spontaneously through amplification of vacuum
fluctuations.\cite{Potasek87} In practice classical input noise
and vacuum fluctuations compete for inducing MI. The experiment
reported here is carried out in the regime where the quantum noise
is dominant.

Elsewhere,\cite{BrainisSTO} we present an unified approach to the
problem of scalar and vector MI based on the \emph{stochastic
nonlinear Schr\"{o}dinger equations} (SNLSE) which generalizes the
work of Ref.~\onlinecite{K91}. This approach is particularly well
suited for numerical simulations in complex situations where
classical noise and vacuum fluctuations act together, where the
pump is depleted, or where the higher order harmonics of MI
appear. In previous work on modulation instability, comparison
between theory and experiment has generally been limited to noting
that the frequency at which the maximum gain of the MI occurs is
correctly predicted. Here we show that there is excellent
agreement between the experimental results and the numerical
integration of the SNLSE. In particular the detailed shape of the
output spectrum can be predicted in detail, even in the case where
higher order harmonics appear (which cannot be predicted by linear
perturbation theory). To our knowledge this is the first time
experimental and theoretical studies of MI are compared in such
detail. A related work is the comparison between theory and
experiment for RF noise measurements reported in
Ref.~\onlinecite{C03}.

\begin{figure}[h]
\begin{center}
\includegraphics[width=84mm]{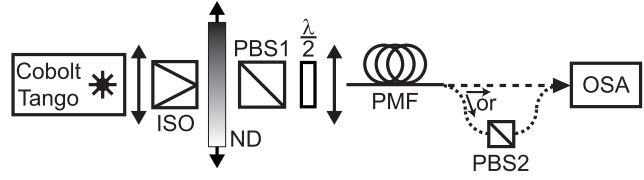}
\caption{Experimental setup. ISO: isolator, ND: variable neutral
density filter, PBS: polarizing beam splitters, PMF: polarization
maintaining fiber, and OSA: optical spectrum analyser.
Amans57847F1.eps} \label{F1}
\end{center}
\end{figure}

The experimental setup is reported in Fig.~\ref{F1}. It consists
of a Q switched laser (Cobolt Tango) that produces pulses at 1536
nm, with a 3.55~ns full-width-at-half-maximum (FWHM) duration
$\tau$ and a 2.5~kHz re\-petition rate $f$. The Gaussian spectral
shape of the laser has been characterized using a Fabry-Perot
interfero\-meter. The measured $0.214$~GHz FWHM spectral width is
slightly larger than expected in the Fourier transform limit.
The pump power is adjusted using variable neutral density filters (ND).
We measured the injected mean power $P_{m}$ at the end of the
fiber. The peak power $P_{0}$ is relied to the mean power
according to the relation:
\begin{equation}
\label{power} P_{0}=2\sqrt{\frac{\ln(2)}{\pi}}\frac{P_m}{
f\tau}=1.06\times10^{5}
P_{m}
,
\end{equation}
A polarizing beam splitter (PBS1) ensures the pump pulse is
linearly polarized. A half-wave plate is used to tune the angle
$\theta$ between the pump polarization direction and the principal
axes of the fiber. A polarizing beam splitter (PBS2) can be used
in order to observe the field components polarized along the fast
or slow axes separately. Lastly, spectra are recorded using an
optical spectral analyser (OSA). In our experiment we use the
Fibercore HB1250P optical fiber. The fiber length $L=51$~m, the
group-velocity dispersion (GVD) parameter $\beta_2=-15$~
ps$^2$~km$^{-1}$ and the group-velocity mismatch parameter
$\Delta\beta_1=286.1$~fs~m$^{-1}$ have been measured by
independent methods (only significant digits have been indicated).
Note that the accuracy on the value of $\beta_2$ is poor compared
to the standards. This is because the interferometric method
\cite{Merritt89} that we used turned out to be difficult to
implement with a birefringent fiber. The group-velocity mismatch
parameter $\Delta\beta_1$ is deduced from the walk-off between
pulses propagating on the principal axes of the fiber. The fiber
length $L$ is deduced from a measurement of the pulse time of
flight. The other important parameters of the fiber, and a more
accurate estimation of $\beta_2$, can be inferred from MI spectral
peaks positions, as explained further.

\begin{figure}[h]
\begin{center}
\includegraphics[width=84mm]{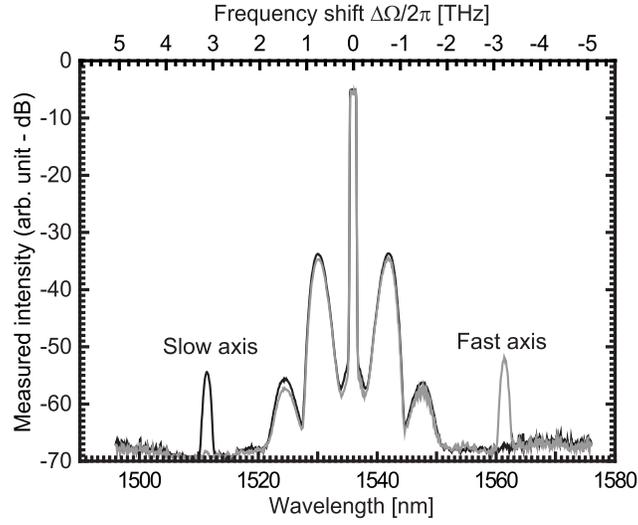}
\caption[]{Spectra of the fast (black curve) and slow (gray curve)
polarization components of the light emerging from the fiber. The
light is injected with a polarization angle $\theta=45^{\circ}$
relative to the principal axes of the fiber. The mean power is
approximatively 1~mW. The resolution bandwidth is 1~nm.
Amans57847F2.eps} \label{F2}
\end{center}
\end{figure}

Fig.~\ref{F2} shows a typical spectrum at the fiber output when
the angle $\theta$ is set to 45 degrees. The fast and slow
polarization components have been separated using PBS2 and their
spectra recorded successively. The plot clearly exhibits two V-MI
peaks at 1511.4~nm and 1561.4~nm that are polarized along the fast
and slow axes respectively. It also shows S-MI peaks at 1530.0~nm
and 1541.9~nm, with first harmonics. In contrast with V-MI, S-MI
is equally generated on both principal axes. By polarizing the
input field along the fast or slow axes, we have observed that
V-MI disappears and that the amplitude of the S-MI peaks increases
dramatically (figure not shown).

\begin{figure}[h]
\centerline{\includegraphics[width=84mm]{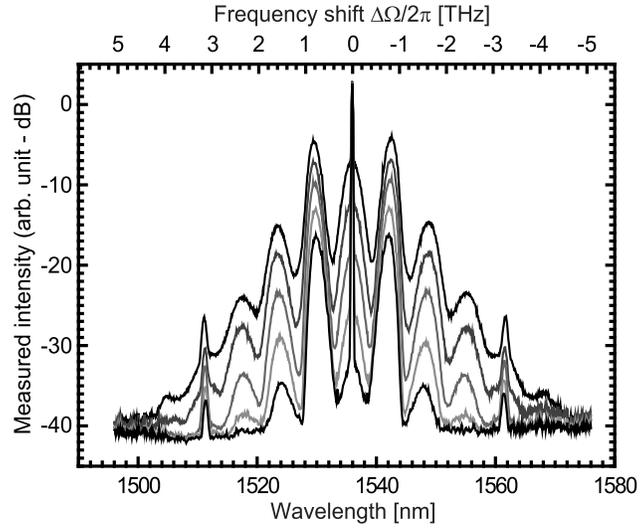}}
\caption{Spectrum of the output field for increasing pump mean
power, respectively $1.065$~mW, $1.105$~mW, $1.145$~mW,
$1.165$~mW, and $1.265$~mW. The light is injected with a
polarization angle $\theta=45^{\circ}$ relative to the principal
axes of the fiber. The resolution bandwidth is 0.1~nm.
Amans57847F3.eps} \label{F3}
\end{figure}

According to linear perturbation analysis, the angular frequency
shifts from the pump of the MI peaks are given by
\begin{eqnarray}
\Delta\Omega_{S-MI}^{2}&\approx&\frac{\gamma
P_0}{|\beta_2|}\left[1-\frac{2}{9}\frac{\gamma P_0}{|\beta_2|}
\left(\frac{|\beta_2|}{\Delta\beta_1}\right)^{2}\right] \label{MIS}\\
\Delta\Omega_{V-MI}^{2}&\approx&\left(\frac{\Delta\beta_1}{|\beta_2|}\right)^{2}\left[1+\frac{2\gamma
P_0}{|\beta_2|}\left(\frac{|\beta_2|}{\Delta\beta_1}\right)^{2}\right]\label{MIV}
\end{eqnarray}
for S-MI and V-MI peaks respectively. Here, $\gamma$ stands for
the Kerr nonlinearity parameter of the fiber. Fig.~\ref{F3} shows
the evolution of the spectrum of light emerging from the fiber
when the pump power is increased. Using Eqs.~(\ref{MIS}) and
(\ref{MIV}), the ratios
$\frac{\Delta\beta_1}{|\beta_2|}=18.740$~(rad)~THz and
$\frac{\gamma}{|\beta_2|}=0.2135$~(rad)~THz$^2$~W$^{-1}$ where
deduced from these measurement. The first ratio and the measured
value of $\Delta\beta_1$ permits to infer that
$\beta_2=-15.27$~ps$^2$~km$^{-1}$, which is compatible with the
independently measured value. From the second ratio, we deduce
that $\gamma=3.26 $~W$^{-1}$~km$^{-1}$.

The exponential growth of the MI peaks and harmonics is clearly
apparent on Fig.~\ref{F3}. From these measurements we deduce that
the ratio between the maximum gain of the V-MI and of the S-MI is
0.67$\pm0.05$, in good agreement with the theoretical value $2/3$.
We also find that the ratio between the maximum gain of the 1st
harmonic and of the S-MI is 1.88$\pm0.15$, in good agreement with
the theoretical value\cite{Hasegawa,Tai86a} of 2.

\begin{figure}[h]
\centerline{\includegraphics[width=84mm]{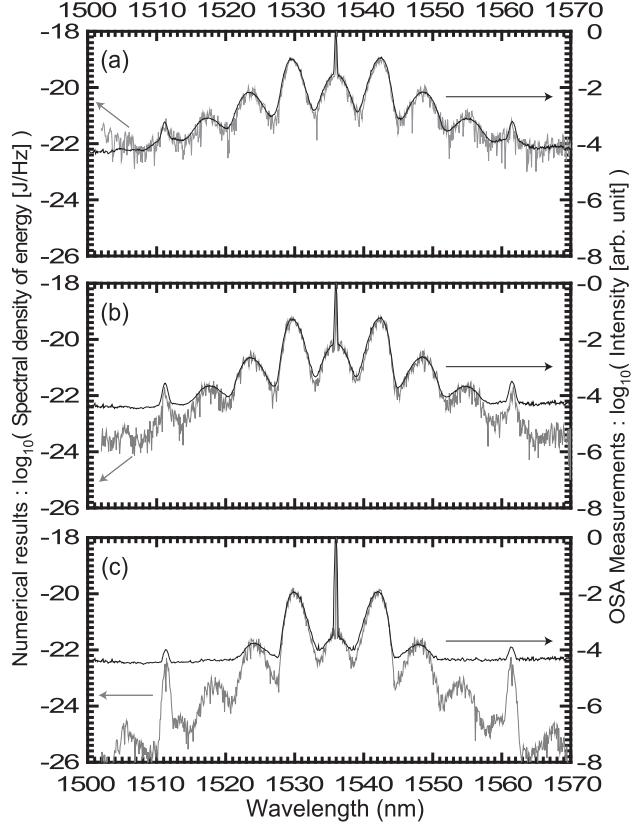}}
\caption{Comparison between experimental spectra (black curves)
and numerical integration of the SNLSE (grey curves). The
numerical results are noisy because only one realization of the
stochastic method has been computed for each curve. The flat parts
on the experimental spectra correspond to the sensibility limit of
the OSA. In the simulations the pump pulse is assumed to be
Fourier transform limited. The simulation parameters are
$\lambda_0=1536$~nm, $\tau=3.55$~ns, $L=51$~m,
$\gamma=3.26$~W$^{-1}$km$^{-1}$, $\beta_2=-15.26$~ps$^2$km$^{-1}$
and $\Delta\beta_1=286.1$~fs m$^{-1}$. The peak powers $P_{0}$
corresponding to simulations (experiments) are (a) 132.5 W (125.6
W), (b) 127.2 W (121.4 W), and (c) 116.6 W (112.9 W). The
numerical results have been convolved with a response function to
take into account the resolution of the OSA. Amans57847F4.eps}
\label{F4}
\end{figure}

We now focus on the quantitative comparison between experimental
spectral amplitudes and those predicted by the SNLSE model for
spontaneous (or vacuum-fluctuations induced) modulation
instabilities. This comparison makes sense because the exact shape
of the spectrum, and in particular the relative intensities of the
modulation instability peaks and harmonics, is very strongly
dependent on the initial noise and pump peak power. Experimental
and computed spectra are plotted together in Fig.~\ref{F4}. In the
simulations we used the parameters deduced from experimental MI
peaks positions (see above), but in order to obtain a good fit we
had to increase the peak pump power by 5\% with respect to that
deduced from the measurements using Eq.~(\ref{power}). We are not
sure of the origin of this discrepancy. It could either be due to
a systematic error in the measured values of $P_m$, to an error in
the experimental measure of $\Delta\beta_1$, to the fact that the
experimental pulses are not exactly Fourier-transform limited, or
to some classical noise photons (for instance due to Raman
scattering in the fiber) that are added to vacuum fluctuations and
slightly speed up the instability. In any case the discrepancy is
 small enough to confidently conclude
that in our experiment the MI is mainly induced by
vacuum-fluctuations. Indeed with this small adjustment the
experimental MI spectra are very well reproduced by numerical
integration of the SNLSE model.

In summary, we report what is to our knowledge the first
experimental observation of spontaneous vector modulation
instability in a highly birefringent silica fiber in the anomalous
dispersion regime. The pump power dependence of the detuning of
both scalar and vector side-bands, as well as their polarizations,
agree with linear perturbation theory when the pump depletion is
small. We have also obtained very good agreement between the
experimental spectra and those obtained by numerical integration
of the SNLSE derived from the quantum theory. This is to our
knowledge the first time that theoretical and experimental spectra
are compared in such quantitative detail. This very good agreement
between the two approaches proves that the modulation instability
that we observed was truly spontaneous, in the sense that it
mainly results from the amplification of vacuum-fluctuations.

\section*{Acknowledgments}
This research is supported by the Interuniversity Attraction Poles
Programme - Belgium Science Policy - under grant V-18. We are also
grateful to Fonds Defay for financial support.


\bibliographystyle{osajnl}

\newpage
{\bf List of Figure Captions}

{Fig. 1.} 
{Experimental setup. ISO: isolator, ND: variable neutral
density filter, PBS: polarizing beam splitters, PMF: polarization
maintaining fiber, and OSA: optical spectrum analyser.}

\bigskip

{Fig. 2.} 
{Spectra of the fast (black curve) and slow (gray curve)
polarization components of the light emerging from the fiber. The
light is injected with a polarization angle $\theta=45^{\circ}$
relative to the principal axes of the fiber. The mean power is
approximatively 1~mW. The resolution bandwidth is 1~nm.}

\bigskip

{Fig. 3.} 
{Spectrum of the output field for increasing pump mean
power, respectively $1.065$~mW, $1.105$~mW, $1.145$~mW,
$1.165$~mW, and $1.265$~mW. The light is injected with a
polarization angle $\theta=45^{\circ}$ relative to the principal
axes of the fiber. The resolution bandwidth is 0.1~nm.}

\bigskip

{Fig. 4.} 
{Comparison between experimental spectra (black curves) and
numerical integration of the SNLSE (grey curves). The numerical
results are noisy because only one realization of the stochastic
method has been computed for each curve. The flat parts on the
experimental spectra correspond to the sensibility limit of the
OSA. In the simulations the pump pulse is assumed to be Fourier
transform limited. The simulation parameters are
$\lambda_0=1536$~nm, $\tau=3.55$~ns, $L=51$~m,
$\gamma=3.26$~W$^{-1}$km$^{-1}$, $\beta_2=-15.26$~ps$^2$km$^{-1}$
and $\Delta\beta_1=286.1$~fs m$^{-1}$. The peak powers $P_{0}$
corresponding to simulations (experiments) are (a) 132.5 W (125.6
W), (b) 127.2 W (121.4 W), and (c) 116.6 W (112.9 W). The
numerical results have been convolved with a response function to
take into account the resolution of the OSA.}

\newpage

\newpage

\newpage

\newpage
\end{document}